\documentclass[aip,cha,reprint,numerical]{revtex4-1}
\usepackage{amsmath}
\usepackage{amssymb}
\usepackage{graphicx}
\usepackage{natbib}		
\usepackage{color}

\def\e{\varepsilon}

\newcommand{\ii}{\mathrm{i}}
\newcommand{\dd}{\mathrm{d}}
\newcommand{\new}[1]{{\color{black} #1}} 

\begin{document}
\title{Inferring collective synchrony observing
spiking of one or several neurons}

\author{Arkady Pikovsky}
\affiliation{Institute of Physics and Astronomy, University of Potsdam,
Karl-Liebknecht-Str. 24/25, 14476 Potsdam-Golm, Germany}
\author{Michael Rosenblum}	 
\affiliation{Institute of Physics and Astronomy, University of Potsdam,
Karl-Liebknecht-Str. 24/25, 14476 Potsdam-Golm, Germany}

\date{\today}

\begin{abstract}
We tackle a quantification of synchrony in a large ensemble of interacting neurons from the observation of spiking events. In a simulation study, we efficiently infer the synchrony level in a neuronal population from a point process reflecting spiking of a small number 
of units and even from a single neuron. We introduce a synchrony measure (order parameter) based on the Bartlett covariance density; this quantity can be easily computed from the recorded point process. This measure is robust concerning missed spikes and, if computed from observing several neurons, does not require spike sorting. We illustrate the approach by modeling populations of spiking or bursting neurons, including the case of sparse synchrony.
\end{abstract}


\keywords{synchrony measure, point process, neuronal population, covariance density}
\maketitle

\section{Introduction}
\label{sec:intro}

 The investigation of neuronal rhythms is a crucial issue in neuroscience, and numerous publications supply evidence of the role of such rhythms in normal and pathological brain 
 functioning~\cite{Singer-93,Buzhaki-Draguhn-04,Buzsaki-06}. 
The emergence of macroscopic rhythmical activity implies a certain level of coordination within a large population of spiking and/or bursting neurons. In terms of nonlinear dynamics, this activity is the collective oscillatory mode arising in a network of active units due to their synchronization. Hence, efficient techniques for quantifying the synchrony level can be helpful in experimental and theoretical studies; this paper aims at developing such a tool for a challenging case when only a few neurons from a large population can be monitored. 

Quantifying collective synchrony is a common task in the physics of complex systems in general, not only in neuroscience. In many situations, the solution is well-known and straightforward. 
\new{So, if the phases of all interacting units are known, the Kuramoto order parameter - a quantity between zero (asynchrony) and one (complete synchrony) - gives the desired answer. An alternative approach is to compute the collective mode's standard deviation; unlike the Kuramoto order parameter, this is not a dimensionless quantity. Thus, one cannot estimate the synchrony level when computing it for a certain network's state. However, this quantity allows one to trace the synchronization transition if one monitors the collective mode's standard deviation and observes its essential increase while varying the degree of interaction between the units.} 
\footnote{\new{For supplementary information, one can compare the standard 
deviation of the collective mode with that of the individual unit's oscillation: in the case of synchrony,
these quantities are of the same order. However, this comparison requires additional measurements.}}

The problem becomes much less trivial if we observe only a small fraction of the population, possibly even one unit. In our recent publication~\cite{pikovsky2024unified}, we suggested a solution to this problem, demonstrating quantification of synchrony from a partial observation under certain conditions. 
However, the approach of Ref.~\cite{pikovsky2024unified} works with oscillators generating smooth signals, while in neuroscience applications, one often deals with spike trains that can be considered as point processes. 
This paper tackles this problem and develops a technique for quantifying synchrony in a large highly interconnected
neuronal network from the recordings of spiking events of one or several units. It is known that such
networks can be treated in the mean-field approximation, i.e., for simplicity, one can assume the global (all-to-all) 
coupling.

The main idea of synchrony quantification is as follows. In a large globally coupled population, an asynchronous state corresponds to a constant (up to small noise due to finite-size effects) mean field acting on the neurons, while a synchronous state corresponds to a regularly (up to small finite-size effects) oscillating mean field.
Let us observe one neuron out of such a population. 
We assume that there is an internal source of noise or chaos, and thus, one uncoupled neuron fires irregularly. 
The same holds if the population is asynchronous because the force from the mean field on the neuron is constant (possibly with small fluctuations). As a result, the spiking of the neuron is purely irregular. Suppose now that the network synchronizes; then, the neuron is driven by the regular collective mode.~\footnote{In terms of neuroscience, the collective mode (mean field) can be associated with the local field potential.} 
This nearly periodic forcing evokes a nearly periodic component in the firing of the observed neuron, and the problem boils down to revealing and quantifying this component from the observation of a noisy process. 
A natural approach is averaging: for the case of smooth oscillations, 
we computed the time average of the squared covariance function, as 
suggested by Wiener's lemma~\cite{wiener1930generalized}, see~\cite{pikovsky2024unified}.
Certainly, the performance is increased if the ensemble averaging complements the time averaging, i.e., the covariance function is computed for the collective mode. However, the technique works with partial observation if a record from one or several units is sufficiently long. This paper extends this idea and suggests a technique appropriate for quantifying periodic components from spike trains, where calculating the usual covariance function is impossible. We will show that instead of the covariance function one can quantify the regular component via the covariance density of the point process~\cite{bartlett1963spectral}.


\section{Materials and Methods}

\subsection{Ensemble models demonstrating synchronization transition}

In this section, we introduce two models of neuronal ensembles demonstrating the transition to collective synchrony with the growth of the coupling strength. In both cases, we assume the interaction of the mean-field type; such coupling is a reasonable model for highly interconnected networks. We exploit both models to generate data sets to illustrate and test the technique developed in Section~\ref{sec:method}. These data sets contain the instants of spikes occurrence, i.e., they are point processes. 

\subsubsection{Noisy spiking and bursting Hindmarsh-Rose neurons}

\begin{figure}
\centering
\includegraphics[width=\columnwidth]{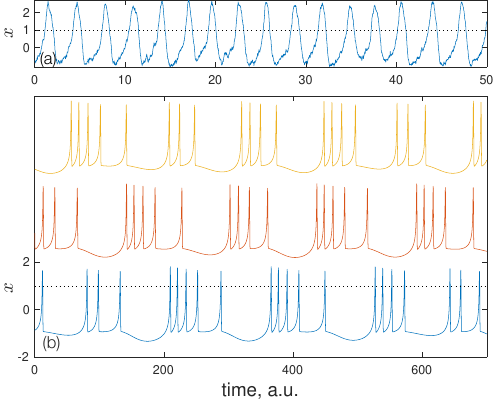}
\caption{The panel (a) illustrates the spiking of a noisy Hindmarsh-Rose neuron, while 
the panel (b) depicts the bursting of HR neurons for $I_n=2.95$, $I_n=3$, and $I_n=3.05$. 
(The second and third curves are shifted upwards for better visibility.) 
From each trace, the point process is generated by taking the instants of the threshold 
crossing from below, the dotted line $x=1$ marks the threshold. 
\new{Notice the difference in time scales in (a) and (b).}}
\label{fig:onehr}
\end{figure}

The Hindmarsh-Rose (HR) model~\cite{Hindmarsh-Rose-84} is a simplified conceptual version of the Hodgkin-Huxley equations~\cite{Hodgkin-Huxley-52}. 
We will exploit its noisy version with global coupling via the mean field $X$:
\begin{equation}
\begin{aligned}
    \dot x_n &=y_n-x_n^3+3x_n^2-z_n+I_n+\sigma\xi_n(t)+\e X\;, \\
    \dot y_n &= 1-5x_n^2-y_n\;, \\
    \dot z_n &=0.006[r(x_n+1.56)-z_n]\;.
\end{aligned}
\label{eq:hr}
\end{equation}
Here $n=1,2,\ldots,N$ is the unit's index; $X=N^{-1}\sum_{k=1}^N x_k$ is the mean field; $\xi_n(t)$ are independent Gaussian white noises with zero mean and unit intensity. Parameters $r$ and $I_n$ define the type of the dynamics; below, we consider two sets of parameters:
\begin{enumerate}
\item $r=1$, $I_n=6$, and $\sigma=0.4$ yield an ensemble of noisy identical spiking neurons;
\item $r=4$, $I_n=2.95+0.1\frac{n-1}{N-1}$, and $\sigma=0$ yield an ensemble of non-identical chaotically bursting units.
\end{enumerate}
In both cases, the time-continuous model is used to generate a sequence of instantaneous 
spikes, or point process, to mimic the observed data. 
For the instants of spike occurrence, we take the moments of threshold $x_n=1$ crossing from below,
see Fig.~\ref{fig:onehr} that illustrates the time dynamics of uncoupled neurons; the height of spikes is ignored.

\begin{figure}[!htb]
\centering
\includegraphics[width=\columnwidth]{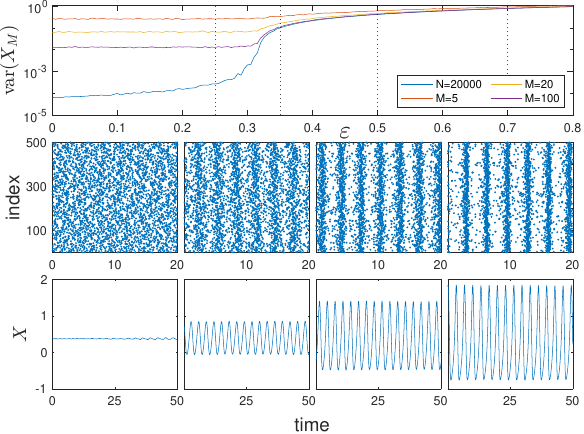}
\caption{The top panel illustrates the synchronization transition in a system of $N=20000$ coupled identical noisy Hindmarsh-Rose neurons, see Eq.~(\ref{eq:hr}); here, we plot the mean field variance as a function of the coupling strength $\e$. The mean field $X_M$ is computed from $M\le N$ neurons. (Notice the logarithmic scale of the vertical axis.)
The plot demonstrates that observation of only $100$ units out of $N$ provides a reliable indication of synchrony.
The middle panels present the raster plots; here, the instants of spiking are marked by a dot
(spikes from only 500 neurons are shown for better visibility). 
The raster plots are given for 
four values of the coupling strength $\e$ marked by dotted vertical lines in the top panel 
($0.25$, $0.35$, $0.5$, $0.7$).
The bottom panels exhibit the corresponding time traces of the mean field $X=X_N$.
}
\label{fig:HRTrans}
\end{figure}

\begin{figure}[!htb]
\centering
\includegraphics[width=\columnwidth]{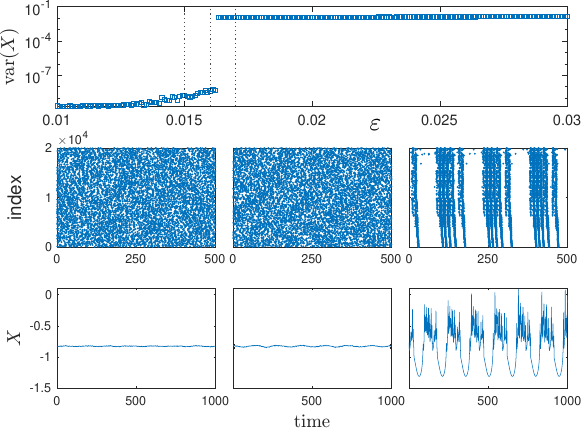}
\caption{Similar to Fig.~\ref{fig:HRTrans} but for bursting nonidentical Hindmarsh-Rose neurons.
Raster plots and mean fields 
are given for $\e=0.015$, $0.016$, and $0.017$ (these values are marked by vertical dotted lines 
in the upper panel); in the raster plots, every 40th neuron is shown.
 Notice that the mean field is irregular in the synchronous state (for $\e=0.017$) but exhibits a strong periodic component. 
}
\label{fig:HRBurstTrans}
\end{figure}

A single neuron ($\e=0$) is producing an irregular, due to noise in the setup 1 or due to chaos in the setup 2, sequence of spikes or bursts, as illustrated in Fig.~\ref{fig:onehr}. A population of neurons, with increase of the coupling strength $\e$, experiences a transition to collective synchrony, as illustrated in Fig.~\ref{fig:HRTrans} for noisy identical spiking neurons and in Fig.~\ref{fig:HRBurstTrans} for nonidentical chaotic bursting neurons. The transition can be determined by following the variance of the mean field $X(t)$. In the thermodynamic limit $N\to\infty$, in the asynchronous regime this mean field is constant, and it is oscillatory beyond the synchronization transition. Using the variance of the mean field as an indicator for synchronization transition has been suggested already in early studies of coupled disordered (noisy or chaotic) systems~\cite{ginzburg1994theory,hansel1996chaos,Pikovsky-Rosenblum-Kurths-96,golomb2001mechanisms}. 
Moreover, one can use the partial mean field $X_M$ averaged not over the whole population, 
but over a subset 
of $M<N$ units~\cite{ginzburg1994theory,hansel1996chaos,golomb2001mechanisms}. 
One expects that in the asynchronous case the variance is $\sim M^{-1}$ due to finite-size effects, 
so for sufficiently large $M$ the transition should be visible. 
Indeed, as can be seen in Fig.~\ref{fig:HRTrans}, the variance of a partial field 
observed from $M=100$ neurons
is almost as good an indicator of synchrony as the variance computed for
$M=N=20000$.
Obviously, the variance of $X_1$, i.e., of the trace of an individual unit, is practically constant and does not provide information about the transition. However, we demonstrate below that we can reveal synchrony by observing only one unit.  

We emphasize that though the raster plots in Fig.~\ref{fig:HRTrans} clearly indicate the emergence 
of order in neuronal firing, revealing this order by observing a single unit is nontrivial. 
For example, the coefficient of variation of interspike intervals barely varies and even increases 
with coupling.

\subsubsection{Brunel-Hakim model for sparse synchrony}

In a series of papers~\cite{brunel1999fast,brunel2008sparsely,ostojic2009synchronization}, Brunel and Hakim suggested and explored a simple model exhibiting nontrivial dynamics where the mean-field frequency is much higher than the firing frequency of individual units; they picked the term sparse synchrony to emphasize this property.
The model comprises noisy leaky integrate-and-fire neurons (LIF) $v_n(t)$ coupled via global fields $X,Y$:
\begin{equation}
\begin{aligned}
    \tau_m\dot v_n &=-v_n + I_0 -\e X +\sigma\sqrt{\tau_m}\xi_n(t) \;, \\
    \tau_d\dot X &= -X+Y\;, \\
    \tau_r\dot Y &=-Y+\frac{\tau_m}{N}\sum \delta\big (t-t_k^{(n)}\big )   \;.
\end{aligned}
\label{eq:BH}
\end{equation}
Here $\xi_n(t)$ are independent Gaussian white noises with zero mean and unit intensity; 
\new{$\tau_m$ is the membrane time constant, and $\tau_r$ and $\tau_d$ are the decay and rise time 
of the postsynaptic current.}
As it is typical for the integrate-and-fire models, a hybrid system \eqref{eq:BH} describes the evolution of the voltage $v_n$ of the neuron $n$ until the voltage achieves a threshold $v_u$; at this moment denoted as $t_k^{(n)}$, the neuron $n$ 
fires an action potential and is instantaneously reset to the level $v_d$. The produced spike contributes to the instantaneous change of the global field $Y(t)$. As in the case of HR neurons, we denote the coupling strength $\e$. \new{The equations for global fields $X,Y$ represent synaptic filtering}.

\begin{figure}[!htb]
\centering
\includegraphics[width=\columnwidth]{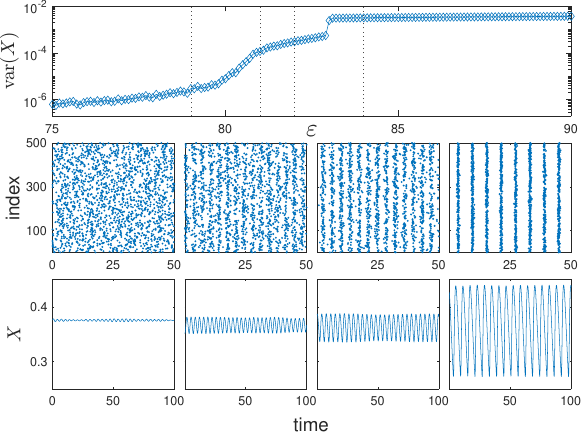}
\includegraphics[width=\columnwidth]{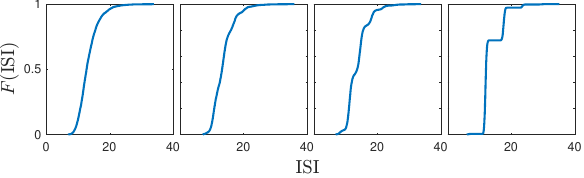}
\caption{The top panel illustrates the synchronization transition for the Brunnel - Hakim 
model Eq.~(\ref{eq:BH}); dotted vertical lines mark the coupling strength values 
$\e=79, 81, 82, 84$ for which the raster plots, mean fields $X$, and cumulative distributions of 
interspike intervals (ISI)  are shown in the second, third, and bottom rows, respectively. 
Population size is $N=10^5$ neurons. 
Notice that the coefficient of ISI variation remains practically constant and does 
not reflect the transition.
}
\label{fig:BHTrans}
\end{figure}

Due to noise, uncoupled noisy LIF neurons produce irregular spike trains. 
With the increase of $\e$, one observes a transition from the disordered state, where fields $X,Y$ are constants (in the thermodynamic limit), to collective synchrony, where the fields $X,Y$ oscillate, 
see Fig.~\ref{fig:BHTrans}.~\footnote{A detailed study shows that the system exhibits bistability and hysteresis in the transition; however, these properties are irrelevant for our study. To avoid ambiguity in the results, in the simulations 
we always use the nearly-asynchronous initial conditions.} 
Remarkably, the period of these oscillations is much smaller than the characteristic interspike interval 
of a single neuron. 
Thus, this regime was termed ``sparsely synchronized neuronal oscillations'' in \cite{brunel2008sparsely}. 
In the simulation, we use the following parameter values: $v_u=5$, $v_d=14$, $\tau_m=5$, $\tau_d=6$, $\tau_r=1$,
$\sigma=0.5$, and $I_0=50$.

\begin{figure}[!htb]
\centering
\includegraphics[width=\columnwidth]{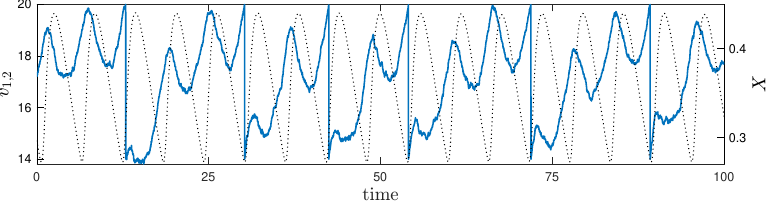}
\caption{Voltage traces of one leaky integrate-and-fire neuron (bold line, left vertical axis) from a population of $N=10^5$ units and of the mean field (dotted line, right axis) for $\e=84$. We see that within the shown time interval, the neuron fires six times. On the one hand, the firing is highly irregular, with the inter-spike intervals
$\approx 17.3,  12.24,   11.58,   17.82,  17.33$. On the other hand, these ISIs are close to the multiples of the 
mean-field period $\approx 5.87$. 
}
\label{fig:BHTime}
\end{figure}

To get more insight into the synchronous dynamics of the Brunel-Hakim system, we depict in Fig.~\ref{fig:BHTime}
the voltage trace of one randomly chosen neuron and the mean field $X$ for $\e=84$. 
Within the shown time interval, the unit fires six times, and one can see that these events occur at approximately the same 
phase of the mean field $X$, which is quite regular. Correspondingly, the interspike intervals (ISI) are approximately 
multiples of the mean-field period $T_X$. (In the shown realization, they are approximately 
$3T_X$, $2T_X$, $2T_X$,  $3T_X$, $3T_X$.) 
This explains the staircase-like cumulative distribution shown in Fig.~\ref{fig:BHTrans}.
This observation means that, though the system is noisy, the neuronal firing is determined mainly by the regular mean field; we denote this regime as strong synchrony.
For a weaker coupling, e.g., for $\e=81$, the mean-field amplitude is small, and, correspondingly, 
the effect of noise is stronger. As a result, the ISI attains different values, not only multiples of $T_X$, and
the corresponding distribution is smooth; we denote this regime as weak synchrony.    

\subsection{Methods for collective dynamics characterization }
\label{sec:method}

This study aims to characterize collective synchrony in an ensemble of spiking or bursting neurons. It assumes that only firing-time measurements are available so that each monitored unit generates a spike train, or a point process.

\subsubsection{Observation of many neurons}
First, we mention a relatively simple case where one observes all or many neurons from the ensemble.
The standard approach presents all observations as a raster plot, where each spike is represented with a dot marker and all available spike trains are superimposed, cf.~Fig.~\ref{fig:HRTrans}. 
Collective synchrony in such a representation appears as a pronounced modulation of the time-dependent density of markers, while in the absence of synchrony, this density is constant.
Thus, the collective synchrony can be inferred by calculating the time-dependent density (instantaneous population firing rate) from the raster plots; a measure of the macroscopic density variation characterizes the synchrony level~\cite{Singer-99,ostojic2009synchronization,Ciba_et_al-18}.

\subsubsection{Observation of one or a few neurons}
\label{sec:oofn}
\begin{figure}
\centering
\includegraphics[width=\columnwidth]{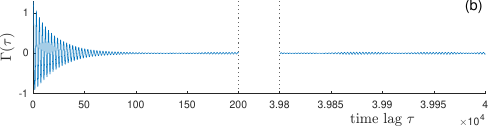}
\includegraphics[width=\columnwidth]{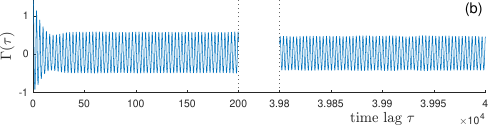}\\[1ex]
\includegraphics[width=\columnwidth]{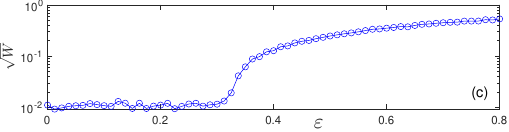}
\caption{Quantification of the ensemble synchrony from observation of only one neuron.
Autocovariance function $\Gamma(\tau)$ computed from the $x$-variable of a 
Hindmarsh-Rose neuron for $\e=0.2$ (a) and  $\e=0.6$ (b). 
Panel (c) presents the square root of the Wiener order parameter $W$ computed from the ACF of one unit vs. the coupling 
strength $\e$; this dependence reveals the synchronization transition, 
cf.~Fig.~\ref{fig:HRTrans}, where many units are used for synchrony quantification via mean-field variance 
(recall that $\sqrt{W}\propto$ variance of the process). 
}
\label{fig:HR_acfs}
\end{figure}

The problem becomes challenging if spike trains are available from only a few units or even from one neuron. A similar problem has been recently addressed in \cite{pikovsky2024unified} for situations where a continuous observable from a unit is available. For the HR system \eqref{eq:hr}, such an observable could be a continuous function of variables $x,y,z$. The method of Ref.~\cite{pikovsky2024unified} relies on the coherence properties of the collective dynamics in the presence of synchrony. The mean field in the synchronous regime is nearly regular, typically periodic (or has a pronounced regular component). Thus, each unit's dynamics contain an internal irregular and a regular component caused by the mean-field driving. So, the problem of synchrony detection is reduced to the extraction and characterization of this regular component. For continuous observables, the proper approach is to calculate the autocovariance function (ACF). The autocovariance tends to zero at a large time lag for a purely irregular signal.
On the other hand, if the signal contains a regular (periodic or quasiperiodic) component, the autocovariance function tends to be a regular, periodic or quasiperiodic, function of the lag at large time lags. According to the Wiener's lemma~\cite{wiener1930generalized}, the average of the squared autocovariance yields a proper quantification of the regular component in the process. Ref.~\cite{pikovsky2024unified} tested this approach with examples of coupled noisy or chaotic oscillators. In particular, for calculating the autocovariance, an observation of only one unit is sufficient (although one can improve the ``signal-to-noise'' ratio if observations from several units are available). Figure~\ref{fig:HR_acfs} illustrates this idea by exploiting the Hindmarsh-Rose model (\ref{eq:hr}) for the case of identical noisy neurons. Here, we show the standard ACFs 
\begin{equation}
\Gamma(\tau)=\frac{1}{T-\tau}\int_0^{T-\tau} (x_1(t)-\bar x_1)(x_1(t+\tau)-\bar x_1)\dd t\;,
\label{eq:stACF}
\end{equation}
estimated from the observation of the $x$-variable of one unit (since they are identical, we choose the first one) for two different values of the coupling strength $\e$; $T$ is the averaging time, and 
$\bar x_1$ denotes time average of the process.
Furthermore, we present the Wiener order parameter $W$, which we estimate as 
\begin{equation}
W(\Gamma)=\frac{1}{\Theta_2-\Theta_1}\int_{\Theta_1}^{\Theta_2} \Gamma^2(\tau)\dd\tau\;,
\label{eq:Wop}
\end{equation}
as a function of $\e$. 
(Since $W$ is proportional to the squared variance of the process, we plot 
$\sqrt{W}$ for a better comparison with Fig.~\ref{fig:HRTrans}.)
We choose the time lags $\Theta_{1,2}$ in such a way that the irregular component due to 
internal noise can be neglected for $\tau>\Theta_1$ so that $\Gamma(\tau)$ can be considered as stationary. Additionally, $\Theta_2-\Theta_1$ must be larger than the characteristic period of $\Gamma(\tau)$.  
Parameters used in Fig.~\ref{fig:HR_acfs} are: $T=10^5$, $\Theta_1=0.03T$, 
$\Theta_2=0.05T$.

\subsubsection{Covariance density for a point process}

The method of  Ref.~\cite{pikovsky2024unified} illustrated in  Section~\ref{sec:oofn} works with continuous-time signals and, hence, cannot be directly applied to spike trains which are point processes.
Here, we present a novel technique, similar to the Wiener's approach, that is suitable for identifying regular components in a point process.

Dealing with a continuous process, one frequently computes the power spectrum and the autocovariance function, which are known to be interrelated by the Fourier transform. The proper corresponding characterizations for point processes are the Bartlett spectrum and the covariance density~\cite{bartlett1963spectral}. Because we apply these concepts to an empirical analysis of particular time series, we write them for a finite observed series of $K$ events $t_1,t_2,\ldots,t_K$. For calculation of the Bartlett spectrum, one formally considers a series of delta-functions at event times, $x(t)=\sum_k \delta(t-t_k)$, and takes its Fourier transform 
\begin{equation}
S(\omega)=K^{-1/2}\sum_{k=1}^K \exp{(\ii \omega t_k)}\;.
\label{eq:ppFt}
\end{equation}
The power spectrum is the average squared absolute value of $S(\omega)$. This latter quantity, which is an analog of a periodogram, reads
\begin{equation}
|S(\omega)|^2=K^{-1}\sum_{k,l=1}^K \exp(\ii\omega(t_k-t_l))=K^{-1}\sum_{m=1}^{K^2} \exp(\ii\omega\tau_m)\;,
\label{eq:bs}
\end{equation}
where we introduced time differences
\begin{equation}
\tau_m=t_k-t_l,\qquad m=1,2,\ldots,K^2\;.
\label{eq:td}
\end{equation}    
These differences are intervals between any pair of spikes; thus, generally they can be also negative. Because of the symmetry $\tau\to-\tau$, below we restrict our attention to positive differences only; furthermore we exclude zero differences by imposing $k>l$.
Equations~(\ref{eq:ppFt},\ref{eq:bs}) show that the power spectrum is the Fourier transform of the averaged effective ``point process'' of the time differences 
$\sum_m \delta(\tau-\tau_m)$, or, in other words, the Fourier transform of the density 
of the set of points $\tau_m$ on the $\tau$-axis;  
Bartlett denoted the density of differences $\tau_m$ between the 
events $t_k$ as the covariance density~\cite{bartlett1963spectral}. 
(Strictly speaking, one also has to subtract the square of the rate of the point process; in our method below, this step, however, is not required.)

Using the analogy with the standard spectral analysis of continuous processes, we conclude that if the Bartlett spectrum contains discrete and continuous components, the same components are present in the covariance density: the continuous component tends for large time lags $\tau$ to a constant, while each discrete component results in the oscillations of the covariance density at large time lags $\tau$. We see that finding regular components in the point process reduces to estimating  oscillating components of the covariance density for sufficiently large time lags $\tau$. Here, we suggest a simple, practical algorithm
for quantification of these regular components.

\subsubsection{Quantification of a regular component in the covariance density}

The first issue is obtaining a non-biased sample of time differences.
Since we deal with a finite-span spike train, the time differences are bounded, $\tau_m\le  T=t_K-t_1$, and the number of pairs of data points $(t_k,t_l)$ having difference $\approx\tau$ decreases for large $\tau$ as $\sim (T-\tau)$. Correspondingly, large time lags $\tau$ are under-represented in the sample if one includes all available pairs $(t_k,t_l)$. A similar issue appears in estimating the autocovariance function; the standard solution there is renormalization by a factor $(T-|\tau|)$,  which yields an unbiased estimation, cf.~Eq.~(\ref{eq:stACF}). We solve the problem of the proper unbiased covariance density estimation by suitably choosing the subset of pairs $(t_k,t_l)$ used to calculate the differences.

\begin{figure}[!htb]
\centering
\includegraphics[width=0.75\columnwidth]{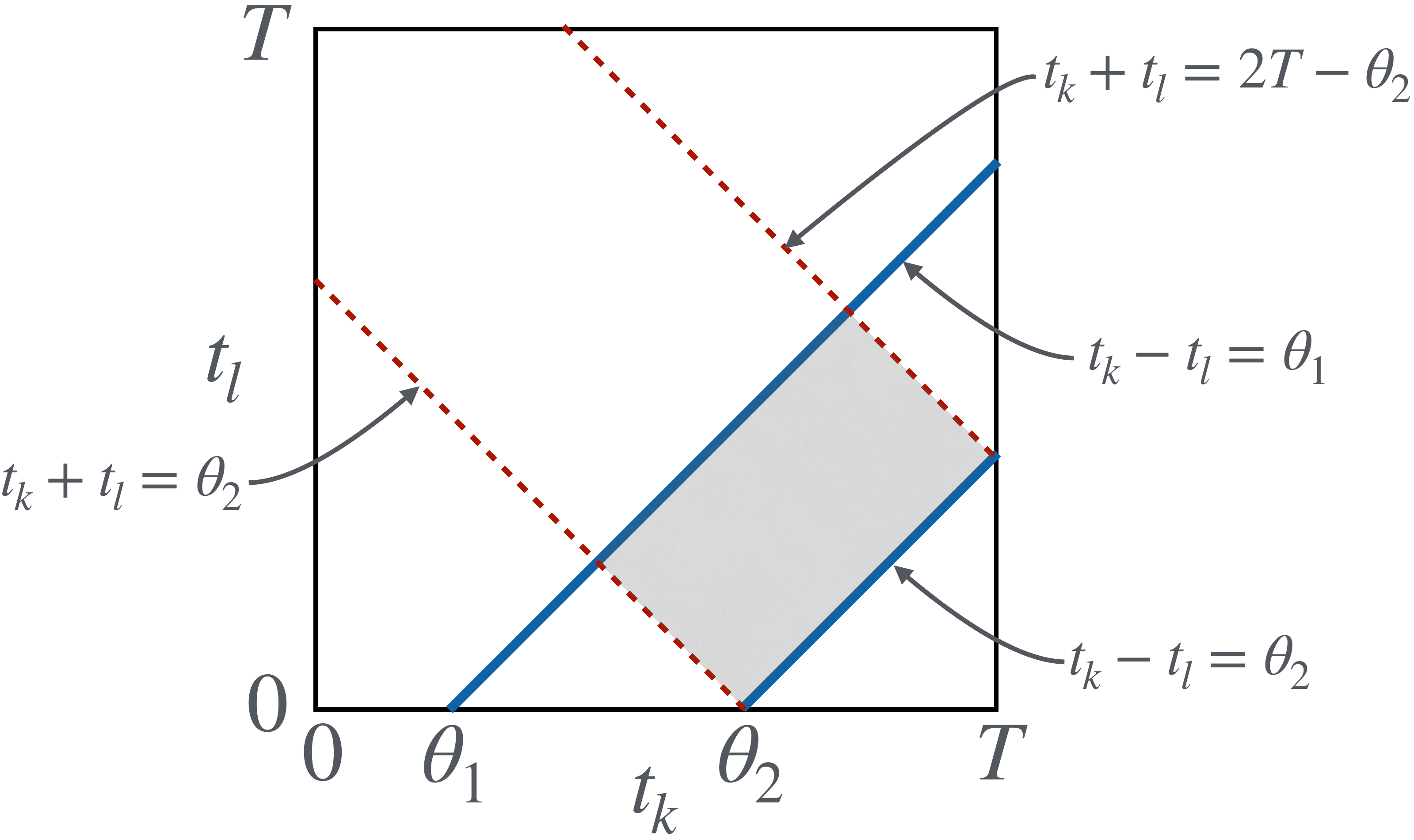}
\caption{Choosing spiking events $t_k$, $t_l$ from the rectangular gray domain 
yields an unbiased covariance density estimation for time differences $\theta_1<t_k-t_l<\theta_2$.
These differences correspond to points in the trapezoidal domain between two blue lines; however, if we use all the points from this domain, the small values $t_k-t_l$ will be overrepresented. }
\label{fig:sketch}
\end{figure}

We illustrate the approach in Fig.~\ref{fig:sketch}. Suppose we want to estimate the covariance density in the interval of time differences $\theta_1<\tau<\theta_2$. Then, selecting pairs of events $(t_k,t_l)$ from the rectangular gray domain determined by the conditions
\begin{equation}
\theta_1<t_k-t_l<\theta_2,\qquad \theta_2<t_k+t_l<2T-\theta_2\;,
\label{eq:int}
\end{equation}
delivers an unbiased sampling, because all the slices of the rectangle with $t_k-t_l=const$ have the same length. Notice that we can also choose $\theta_1=0$. 
Thus, starting with the original point process $\{t_k\}$ and picking up all the events 
according to \eqref{eq:int}, we obtain a subset of length $L$ of time differences $\tau_m$, 
$m=1\ldots,L$. The size of the sample can be estimated as $L\approx K(K-1)(\theta_2-\theta_1)(T-\theta_2)T^{-2}$.

Next, we need to check whether the density of the sampled points $\tau_m$ is uniform. As usual in the testing of sampled distributions, it is convenient to work not with the covariance density itself but with the corresponding cumulative distribution, which is the integral of the density. For a similar approach to characterizing a distribution of interspike intervals see Ref.~\cite{Pu-20}. Thus, we define the empirical cumulative covariance distribution function as
\begin{equation}
F(\tau)=\frac{1}{L}\sum_{m=1}^L H(\tau-\tau_{(m)})\;,
\label{eq:f}
\end{equation}
where $H(x)$ is the Heaviside step function.  
Here $\tau_{(m)}$ is the order statistics of the set of time differences, i.e., the $m$-th smallest value of $\tau$; one obtains it by sorting the array in ascending order. 
The inverse of the distribution function
is the quantile function $\tau_{(1)}\leq Q(p)\leq\tau_{(L)}$, $0\leq p\leq 1$.
For this quantile function, we can define the Kantorovich-Rubinstein-Wasserstein (KRW) distance~\cite{kantorovich1958space,vaserstein1969markov} (see \cite{bobkov2019one} for a general introduction of the application of the KRW distance to empirical measures) to the uniform quantile function $Q^u(p)$ as
\begin{equation}
\rho_1(Q,Q^u)=\int_0^1 |Q(p)-Q^u(p)|\dd p\;.
\label{eq:q}
\end{equation}
One possible choice for $Q^u$ would be $Q^u(p)=\tau_{(1)}+p(\tau_{(L)}-\tau_{(1)})$. However, such a choice is sensitive to boundary effects. Thus, we suggest to apply a linear fit to the set $\tau_{(m)}$, 
i.e., to approximate it as $\tau_{(m)}\approx a+bm$ (for a standard procedure for the linear fit computation, see, e.g., in \cite{Press-92}). 
Fitting corresponds to taking the uniform quantile function as $Q^u(p)=a+bLp$. Substitution of this function  and the empirical cumulative covariance distribution function \eqref{eq:f} \eqref{eq:q} 
leads, after representing the integral as a sum, to the following simple formula for the KRW distance
\begin{equation}
\rho_1(Q,Q^u)\approx L^{-1}\sum_{m=1}^L |\tau_{(m)}-a-b m|\;.
\label{eq:q1}
\end{equation}

For the visualization of the covariance density, it is convenient to introduce an additional notation 
for the function under the sum in \eqref{eq:q1}: 
\begin{equation}
C(\tau_{(m)})=\tau_{(m)}-a-bm\;.
\label{eq:cd}
\end{equation}
In fact, this is not an empirical covariance density but rather an integral of it, with the linear trend subtracted. Nevertheless, it has the same properties as the covariance density: for a purely random set of point events, it vanishes for large $\tau$, but if there is a regular component in the point process, function $C(\tau)$ will for large $\tau$ demonstrate the corresponding regularity. We will call it the Empirical Cumulative  Covariance Distribution Function (ECCDF). Note that subtraction of the squared rate of the point process is not required because the linear fit automatically eliminates it.

Summarizing, we suggest to characterize the level of regularity in a time series of the point process by calculating the KRW distance according to
\begin{equation}
D=L^{-1}\sum_{m=1}^{L} |C(\tau_{(m)})|\;.
\label{eq:krw}
\end{equation}

\begin{figure}[!htb]
\centering
\includegraphics[width=0.75\columnwidth]{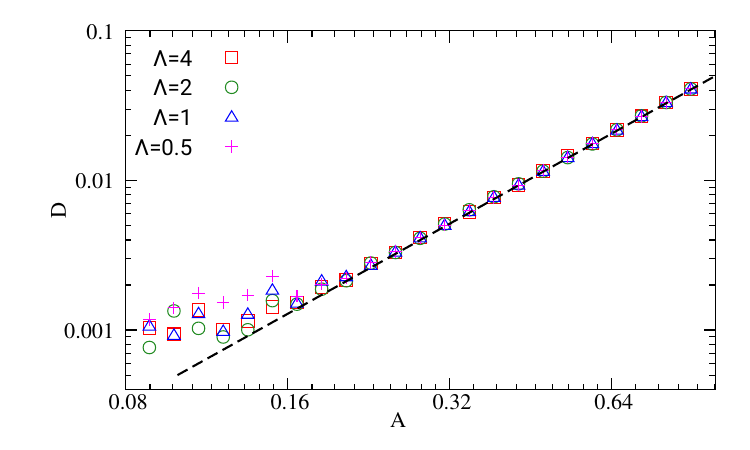}
\caption{KRW distance $D$, see \eqref{eq:krw}, vs. parameter $A$ of the periodically modulated Poisson process, for different values of $\Lambda$. The dashed line has slope $2$.}
\label{fig:poi}
\end{figure}

To illustrate this measure of regularity in a point process, we applied it to a synthetic data set generated as a Poisson process with independent time intervals between the points and time-dependent instantaneous rate $\lambda(t)=\Lambda(1+A\sin(2\pi t))$. Thus, the regular component has a unit period, and its amplitude is $\sim A$. 
We calculated the KRW distance according to Eq.~\eqref{eq:krw} for $L= 2\cdot 10^7$ with parameter values $\theta_1=0,\;\theta_2=20$ (see Eq.~\eqref{eq:int}). 
The results presented in Fig.~\ref{fig:poi} show that the regular component for the parameters chosen can be reliably estimated for $A\gtrsim 0.1$. Notably, as one expects for a covariance measure, $D\propto A^2$.

Since $D$ quantifies a regular component in the dynamics, it serves as an order parameter for the transition to synchrony.~\footnote{One can also consider generalized KRW distances by changing the definition of the norm in \eqref{eq:krw}; in general, one could calculate a $q$-norm as 
$(\sum_{ m} |C(\tau_{(m)})|^q)^{1/q}$. However, only for $q=1$ the simple representation \eqref{eq:q1} is valid.} Of course, this parameter does not vanish in the asynchronous case for finite samples, but for large $L$ its value can be relatively small~\cite{del1999central}. 
We emphasize the similarity between the order parameters $D$ (Eq.~\eqref{eq:krw}) and $W$ (Eq.~\eqref{eq:Wop}): both calculations involve two constants. However, while in the case of continuous signals, constants $\Theta_{1,2}$ explicitly enter Eq.~\eqref{eq:Wop},  for the point processes, constants $\theta_{1,2}$ appear implicitly in the selection of the time differences $\tau_m$ according to the condition \eqref{eq:int}.  

Finally, we discuss the choice of constants in the suggested technique. We assume that the sequence of times $\{t_k\}$ is given; indeed, the approach works better if the time series is longer. The only two constants to choose are $\theta_1,\theta_2$ in \eqref{eq:int}. Constant $\theta_1$ should be larger than the characteristic time of correlations' decay in the process's irregular component. Thus, it is advisable to start with $\theta_1=0$ and plot the values $C(\tau)$ in some interval $0<\tau<\theta_2$. This graph looks roughly like an autocovariance function for a continuous process: it decays initially, and from some values of $\tau$ it either fluctuates around zero or has a regular (typically periodic) tail. According to this picture, for calculating the KRW order parameter $D$, $\theta_1$ should be chosen larger than the initial correlation decay time. The value of $\theta_2$ should not be too large because the statistics will be poor according to the restriction \eqref{eq:int}. On the other hand, the interval $\theta_2-\theta_1$ should include at least several periods of the characteristic variations of the covariance density in the regular tail. 

For synchronization detection, the finite size $N$ of the original ensemble also plays a role. Indeed, the mean fields are purely regular only in the thermodynamic limit of an infinite number of elements ($N\to\infty$ in Eqs.~\eqref{eq:hr},\eqref{eq:BH}).  For large but finite populations, there are finite-size fluctuations that lead to a slow loss of coherence of the mean-field dynamics. As a result, periodic variations of the covariance density discussed above decay on a large time scale $T_{cor}$. This decay does not allow for very large values of the constant $\theta_2$, which bounds the range of time lags; one should preferably take $\theta_2\ll T_{cor}$, or, if the time scale $T_{cor}$ is relatively small, to take $\theta_2\lesssim T_{cor}$.

\begin{figure}[!htb]
\centering
\includegraphics[width=0.75\columnwidth]{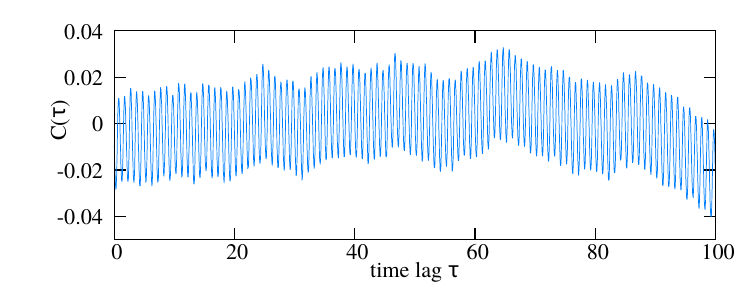}
\caption{The ECCDF for the modulated Poisson process plotted over a large interval of time lags exhibits a slow diffusion-like behavior due to finite-size noise.}
\label{fig:diff}
\end{figure}

There is also a technical reason to keep the value $\theta_2$ relatively small. We work with a finite number of events, resulting in a statistical noise in the empirical covariance density. The cumulative distribution function \eqref{eq:cd} that we analyze contains the integral of this noise. Thus, the finite-size (in the sense of finite length of time series) noise results in a slow diffusion-like behavior at large intervals of the time lag $\tau$. This diffusion-like trend is superposed with a regular component, as is illustrated in Fig.~\ref{fig:diff}. On very large time lag intervals, the contribution of the diffusion part to the KRW distance $D$ (see Eq.~\eqref{eq:krw}) can be significant. Thus, it is advisable to take the time span $\theta_2-\theta_1$ less than the characteristic diffusion time. For example, for the data in Fig.~\ref{fig:diff}, one can use $0\leq \tau\leq 20$. Another way is to eliminate the effect of diffusion by virtue of a detrending procedure, but this will bring extra parameters to the method. 

Finally, we mention that although we focused above on a time series from one neuron, the same method works if we include contributions from several or many units in the point process under consideration. The reason is that under the assumption of global coupling, the same regular mean field acts on different ensemble units, and they share the same regularity in the spiking events. The covariance density, in this case, is, in fact, a mixture of self-covariance (if the difference of spike instants from one unit is taken) and cross-covariance (if one takes the difference of spike instants from two different units). Both the self-covariance and the cross-covariance at large time lags reveal the same regular mean field; thus, ``mixing'' of entries from different units does not prevent detection of this mean field. We provide examples of quantification of such ``mixed'' point processes below.

\section{Results}
We first illustrate the main idea of our approach by exploiting the simple model of spiking HR neurons, 
and then proceed with the more complicated cases of bursting and sparse synchrony.

\subsection{Hindmarsh-Rose neurons}
\subsubsection{Identical noisy spiking neurons}

\begin{figure}[!htb]
\centering
\includegraphics[width=\columnwidth]{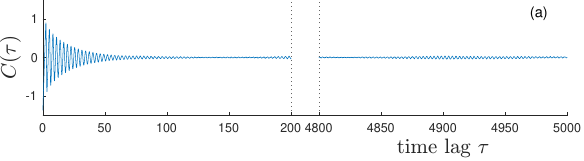}\\
\includegraphics[width=\columnwidth]{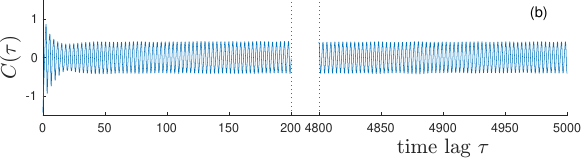}\\[1ex]
\caption{ECCDFs for the sub-threshold $\e=0.2$ (panel a) and super-threshold 
coupling $\e=0.6$ (panel b). 
}
\label{fig:HR_cdfs}
\end{figure}

\begin{figure}[!htb]
\centering
\includegraphics[width=\columnwidth]{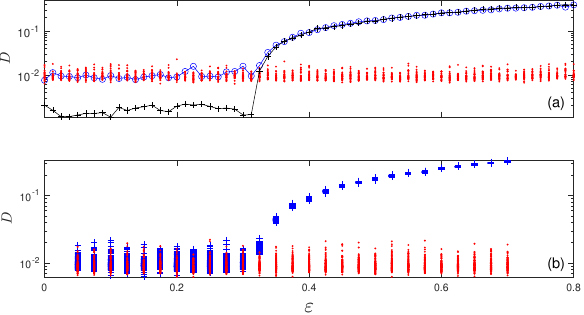}
\caption{
Panel (a) presents the order parameter $D$, see Eq.~(\ref{eq:krw}), computed for 
$\theta_1=2000$, $\theta_2=3000$. Blue circles show the values obtained from observing a single neuron, 
while black pluses correspond to observing a hundred of units. 
Red dots show the order parameter obtained from surrogate point processes; 
see text for details.
Panel (b): here, the blue pluses depict $D$ values computed computed from different single neurons; 
for each value of the coupling, $D$ was computed hundred times. 
Surrogates obtained from each of used point processes are shown by red dots. We see, that in the synchronous state
the $D$ value is almost independent of the chosen neuron. 
}
\label{fig:HR_Dop}
\end{figure}

In the first test, we consider an ensemble of noisy identical neurons as illustrated in Fig.~\ref{fig:HRTrans}, and present the results in Fig.~\ref{fig:HR_cdfs}. 
Here, we show the ECCDFs for weak, $\e=0.2$, panel (a), and strong, $\e=0.6$, panel (b), coupling.
The main message here is that for weak coupling, ECCDF decays and fluctuates around zero, 
while in the second case, the ECCDF's envelope tends to be constant due to the decay of
the continuous spectral component, while the discrete component 
remains.~\footnote{To eliminate the diffusion-like behavior due to the finite-size effect, we remove the linear fit 
separately for each shown interval.}  
The latter can be used for synchrony quantification using  Eq.~(\ref{eq:krw}).
We compute the KRW order parameter $D$ for different values of the coupling 
strength $\varepsilon$; for each $\e$ we do it for two point processes. 
The first process contains spikes from one neuron within an observation interval of length $T$ 
(since the neurons are identical, we take the first one). 
The second point process is constructed from the spikes of hundred neurons observed within the 
interval $T/100$. By construction, the number of spikes in both processes is approximately 
the same (about 35000 spikes; the exact number depends on $\e$). 
The results in Fig.~\ref{fig:HR_Dop}(a) demonstrate that the order parameter values computed from these two 
processes practically coincide in the synchronous state, though the contrast between asynchronous and synchronous states
is higher if 100 neurons are observed. 
The dependencies $D(\e)$ shall be compared with the similar curve $\sqrt{W}(\e)$
in Fig.~\ref{fig:HR_acfs}c. We see that synchrony quantification from point processes works as successfully as quantification from a continuous signal, though much less information is used in the former case.

Furthermore, Fig.~\ref{fig:HR_Dop}(a) shows the results of a simple surrogate test. 
In this test, starting with the process $\{t_k\}$, we construct new point processes $\{t^{(s)}_k\}$. 
For this purpose, we compute the interspike intervals $d_k=t_{k+1}-t_k$, obtain a sequence of new intervals  $d^{(s)}_k$ 
by a random permutation (reshuffling) of $d_k$, and take $t^{(s)}_1=t_1$,  
$t^{(s)}_k=t^{(s)}_{k-1}+d^{(s)}_{k-1}$ for $k>1$. Finally, we compute the KRW order parameter 
from  $\{t^{(s)}_k\}$ and show the obtained value by a red dot. 
For each value of the coupling strength, we perform 25 surrogate tests (we use observations of a single unit). 
We see, that for $\e\ge 0.35$ the values obtained from the ``true'' point process clearly exceed
the surrogate values and thus reliably indicate the presence of synchrony.

Next, we illustrate the precision and robustness of our algorithm by choosing observations of different units. 
Specifically, we compute $D$ a hundred times for a fixed coupling value, each time from a different single unit.
The results in Fig.~\ref{fig:HR_Dop}(b) show that the precision increases with the coupling: 
beyond the synchronization transition, different units provide nearly coinciding results

\begin{figure}[!htb]
\centering
\includegraphics[width=0.6\columnwidth]{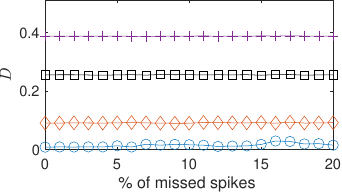}
\caption{Robustness of the synchrony quantification in case of imperfect measurement missing some spikes, for $\e=0.2$ (circles), $\e=0.4$ (diamonds), $\e=0.6$ (squares), and $\e=0.8$ (pluses).  
}
\label{fig:HR_robustness}
\end{figure}

Finally, we check the technique's robustness concerning missed spikes. 
With this test, we imitate the situation when the measurement is imperfect and does 
not record some spikes. 
We start with the point process $t_k$, compute the order parameter $D$, eliminate some randomly 
chosen events, and repeat the computation of $D$. 
Figure~\ref{fig:HR_robustness} presents the dependence of the estimated order 
parameter on the percentage of missed spikes for four values of the coupling strength. 
We see that the synchrony quantification is not robust only for weak coupling when the periodic component is also weak. However, even in this case, the omission of $15\%$ of spikes yields nearly constant results. 

\subsubsection{Non-identical chaotic bursting neurons}

\begin{figure}[!htb]
\centering
\includegraphics[width=\columnwidth]{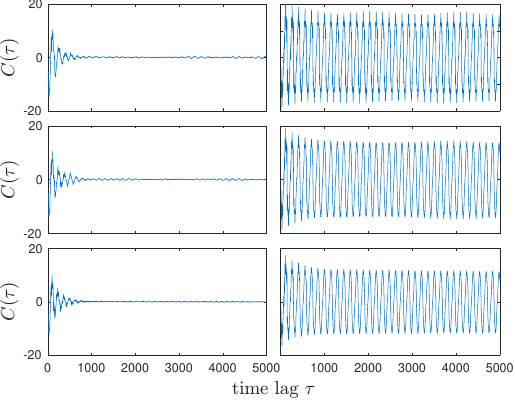}
\caption{ECCDFs for sub-threshold coupling $\e=0.015$ (left column) and for super-threshold 
coupling, $\e=0.017$, for three neurons (see text).
}
\label{fig:HRburst_cdfs}
\end{figure}

In the next test, we take the nonidentical bursting HR units. The challenge here is that the mean field
of this ensemble is obviously irregular (see Fig.~\ref{fig:HRBurstTrans}), though has a 
pronounced periodic component.
We choose two values of the coupling, 
$\e=0.015$ and $\e=0.017$, i.e., below and above the synchronization threshold. 
Figure~\ref{fig:HRburst_cdfs} demonstrates that our approach works for the bursting data. 
Here, we show the results for three neurons, with parameter values
$I_n=2.95$ (bottom panels), $I_n=3$ (middle panels), and $I_n=3.05$ (top panels), cf.~Fig.~\ref{fig:onehr}. 
The corresponding values of the order parameter are:
\begin{enumerate}
\item Sub-threshold coupling, $\e=0.015$: $D=0.11, 0.10, 0.09$.
\item Super-threshold coupling, $\e=0.017$: $D=9.39, 8.83, 7.35$.
\end{enumerate}


\subsection{Brunel-Hakim model}

\begin{figure}[!htb]
\centering
\includegraphics[width=\columnwidth]{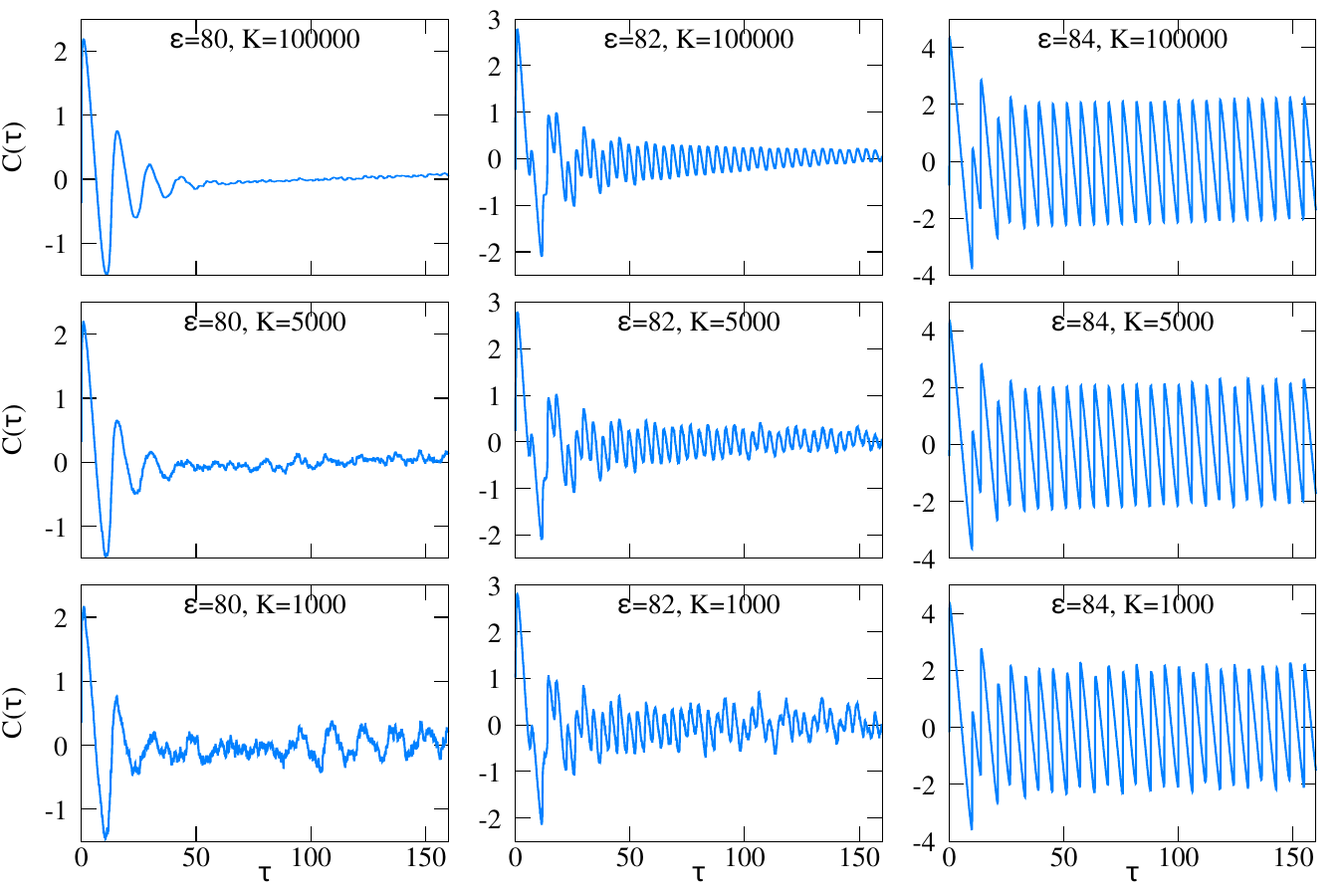}
\caption{Estimations of the ECCDFs for the BH system \eqref{eq:BH} for different 
values of the coupling strength $\e=80$ (asynchrony), $\e=82$ (weak synchrony), and $\e=84$ (strong synchrony). Together with the data with a maximal time series length of $K=10^5$ spikes (top row), we show the results for shorter time series $K=5000$ and $K=1000$ spikes. The latter ECCDFs are much more noisy, except for the regime of strong synchrony. }
\label{fig:bhcd}
\end{figure}

In Fig.~\ref{fig:bhcd}, we report on applying the method to the Brunel-Hakim model \eqref{eq:BH}.
Empirical Cumulative Covariance Distribution Functions are shown for several coupling parameter values $\e$, for an ensemble of $N=100000$ neurons. A time series of $K=10^5$ spikes from a single unit was used in all cases. One can see that the calculations of the ECCDF for a single neuron reliably reveal the synchronization transition. The initial decay of correlations due to irregularity of a single neuron happens within time interval $\tau\lesssim 60$. On the other hand, the decay of correlations of the regular component happens on a much longer time scale. In this example, an appropriate choice of the values $\theta_{1,2}$ could be $\theta_1=60$, $\theta_2=120$.  
Furthermore, in Fig.~\ref{fig:bhcd}, we show how the quality of the covariance density estimation depends on the length of the sample. One can see that although the transition can be traced also for short time series with $K=5000$ and $K=1000$, the quality of the covariance density estimation becomes really poor at this length of the point process.

\begin{figure}[!htb]
\centering
\includegraphics[width=0.8\columnwidth]{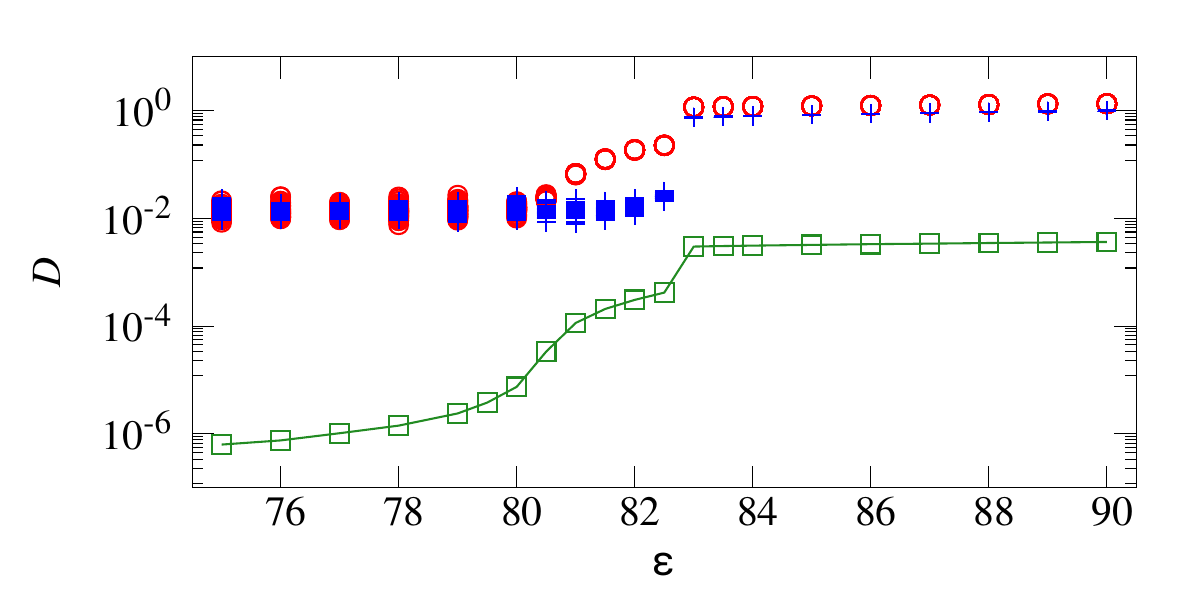}
\caption{Order parameter $D$ vs. coupling strength $\e$ for the BH model. Here the time series of $K=10^5$ spikes was used. Red circles: data for 50 different neurons. Blue pluses: the same time series but randomly reshuffled. For comparison, we also show the variance of the mean field $X(t)$ with green squares. }
\label{fig:bhcd1}
\end{figure}

We report the calculations of the order parameter $D$ in dependence on the coupling constant $\e$ in Fig.~\ref{fig:bhcd1}. Here, for comparison, also the variance of the field $X(t)$ is depicted. For each value of $\e$ we  calculated $D$ from spike trains produced by 50 different neurons, all these data are plotted with red circles. While in the asynchronous state $\e\lesssim 80.5$ there is a large diversity of the values of $D$, in the synchronous states these values nearly coincide so that the markers overlap. Additionally, we report the calculations of $D$ for the randomly shuffled time series (blue pluses). In the regime of weak synchrony ($\e\leq 82.5$), the values of $D$ for the shuffled data are much less than for the original  time series, thus this method of verification works well. However, in the regime of strong synchrony ($\e\geq 83$) also reshuffled data beget large vaues of the order parameter $D$. This is related to a highly peaked distribution of the interspike intervals (see Fig.~\ref{fig:BHTrans}): almost all the ISI are multiples of the period of the mean field, and this property leads to a large ECCDF even after reshuffling. Hence, in this case, the surrogate test is not really required:  a simple analysis of the interspike intervals distribution complements our analysis and supports the conclusion about the presence of synchrony.

\begin{figure}[!htb]
\centering
\includegraphics[width=\columnwidth]{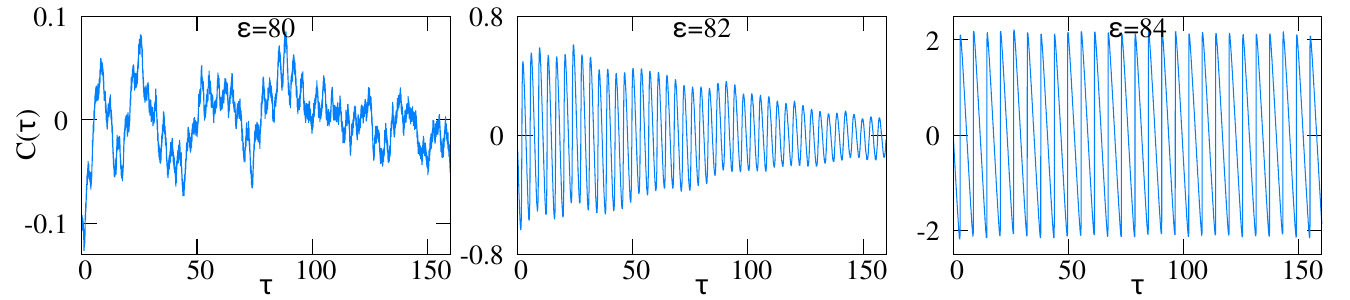}
\caption{Estimations of ECCDFs for a randomized subset of events for the BH model.}
\label{fig:bhcd2}
\end{figure}

Above, we discussed the construction of the covariance density from the point process from one unit or a combination of observations from several units. With Fig.~\ref{fig:bhcd2}, we illustrate the construction of the ECCDF from \textit{randomized} many-unit observations. Namely, we first recorded point processes from 50 units, totaling $5\cdot 10^6$ events ($\approx 10^5$ for each unit). Then, we have chosen randomly (using uniform distribution) $10^5$ events from this set. This means that there are $\approx 2000$ events from each unit. We constructed the ECCDF from the time differences of the selected events as described above in Section~\ref{sec:method}. Because the data include events from many units, it is better to speak on ``cross-covariance density'' in this case, which is analogous to cross-correlations. As the results presented in Fig.~\ref{fig:bhcd2} show, the contrast between the synchronous regime with a large periodic component and the asynchronous state with a small irregular component is evident.

We emphasize the practical advantage of a multi-unit observation. For simplicity, we used global coupling in our test examples. In reality, the neurons are coupled to the local field potential with a different strength. Thus, the one-unit analysis generally depends on the unit. Multi-unit measurement is equivalent to averaging over a (small) subpopulation and, therefore, provides a more reliable estimation of the population synchrony level.

\section{Discussion and conclusion}
We demonstrated that an appropriate processing of one or several units' activity reveals a macroscopic rhythm in a neuronal population.  This result aligns with the findings that a small subset of neuronal populations is involved in the visual perception of complex images \citep{QuianQuiroga-05} or conveys information sufficient for neuroprosthetics 
purposes \citep{Willett_et_al-23}.
The developed technique quantifies the level of synchrony in the neuronal population from the observed spiking events obtained via threshold-crossing. 
The height of the spikes is not accounted for, so the input data are point processes.
The analysis of the covariance density is mathematically equivalent to the analysis of the Bartlett spectral measure.
Indeed, the transition to synchrony is reflected by the appearance of a discrete spectral component (in the limit $N\to\infty$) 
what corresponds to non-decaying $C(\tau)$. Thus, the measure $D$ quantifies the discrete spectral component, and this computation is
analogous to Wiener's lemma, which yields a similar quantity for continuous processes from the autocovariance function.

The simple and computationally inexpensive algorithm provides a single number, which we denote as the KRW 
distance $D$. This order parameter successfully distinguishes synchronous and asynchronous states.
The drawback of this measure is that it is not normalized; hence, a single computation
does not say whether $D$ is large (synchrony) or small (asynchrony). In this case, one has to plot the 
empirical cumulative covariance distribution function $C(\tau)$ and inspect it visually. 
Another ad hoc solution is to rely on the surrogate test. 
However, if observations of synchronous and asynchronous states 
are given, the corresponding values of $D$ differ by more than one order of magnitude.
A clear advantage of the approach is its stability regarding imperfect measurement -- missing about 10 or 15 percent of 
spikes do not affect the result.  
Another beneficial feature is that the number of neurons contributing to the observed point process is irrelevant, and spike sorting is not needed.
The suggested technique works equally well with the data from one neuron and a mixed point process containing spikes from many neurons.
Actually, in the last case, the performance is even better (though the number of these neurons remains much smaller than the population size). These features make the developed approach an efficient tool for experimental and model studies. 
\new{This paper presented examples where we considered statistically stationary situations only. Stationarity allowed for analyzing a rather long time series, for which regularity detection in spike trains works reliably. In a non-stationary case, one has to perform the analysis within a finite time window, i.e. with a rather small time series. As we have demonstrated in Fig.~\ref{fig:bhcd}, for a short sequence of spikes the ECCDF $C(\tau)$ becomes rather noisy. The data of Fig.~\ref{fig:bhcd} suggest that at least several hundred spikes are necessary for regularity detection. A more detailed statistical analysis is a subject of future studies.}

\new{This paper focuses on characterizing regularity in the observed point process or several point processes. In some cases, continuous-time measurements of the Local Field Potentials (LFP) 
can be performed simultaneously with the registration of the spiking activity.
If LFP observations are available, one can use them to characterize the regularity of the neuronal population using well-established methods for continuous processes. The correspondence between the information provided by the LFP analysis and our technique needs further clarification. It is known that the variation of the 
LFP reflects spiking multi-unit activity, but their interrelationship is complicated and depends on the 
frequency range~\citep{Telenczuk-22}. Since the LFP-spike interrelation is very sensitive to neuronal correlations~\citep{Telenczuk-22}, we expect that for strong synchrony, the LFP analysis 
and point process processing provide similar conclusions, while for low and intermediate synchrony levels, 
our technique may be more informative. Additionally, the LFP is a broad-band signal, and the results of 
its spectral-based quantification depend on the choice of the frequencies of interest and preprocessing techniques, such as filtration. We also mention that point process analysis admits a simple surrogate test for significance (reshuffling of inter-spike intervals), while the corresponding tests for continuous-time processes are problematic.
}

\new{Finally, while the regularity of spiking is usually associated with synchrony, the sources of the synchrony can potentially be different. In the examples explored in this paper, synchrony appears due to the interaction of the neurons within the population. However, a similar state can appear due to regular external action on the neurons from other brain regions or external fields. Clearly, the method of regularity detection proposed in this paper cannot resolve the source of regularity; however, it can be a part of a more comprehensive analysis of larger brain areas.}

We thank A. Neiman and V. Nikulin for useful comments.

%
%
%
%
%


\end{document}